\documentclass{article}
\usepackage[top=1in,left=1in,bottom=1in,right=1in, paperheight=11in, paperwidth=8.5in]{geometry}
\usepackage[utf8]{inputenc}
\usepackage{natbib}
\usepackage{setspace}
\usepackage{bm}
\usepackage{xcolor}
\usepackage{amsmath}
\title{Advancing Microdata Privacy Protection: \\ A Review of Synthetic Data}
\author{Jingchen Hu and Claire McKay Bowen}
\date{Vassar College and Urban Institute\\
\today}

\begin{document}

\maketitle

\begin{abstract}
Synthetic data generation is a powerful tool for privacy protection when considering public release of record-level data files. Initially proposed about three decades ago, it has generated significant research and application interest. To meet the pressing demand of data privacy protection in a variety of contexts, the field needs more researchers and practitioners. This review provides a comprehensive introduction to synthetic data, including technical details of their generation and evaluation. Our review also addresses the challenges and limitations of synthetic data, discusses practical applications, and provides thoughts for future work.    
\end{abstract}

% {\textcolor{red}{Items to pay attention to when revising:}}

% \begin{itemize}
%     \item \textcolor{red}{Tenses used for references. Monika is using current tense.}
%     \item \textcolor{red}{The use of synthetic data as singular (referring to the method) and as plural (referring to the actual data) can be confusing and we will need to make the distinctions and be consistent.}
% \end{itemize}

\section{Synthetic data generation is an unreal reality}
\label{sec:intro}
% ``Introduce your topic in ~2 paragraphs, ~750 words."
Synthetic data generation is an unreal reality. As the phrase suggests, the data are not real but is a popular method for protecting confidential data.\footnote{We distinguish two types of data: original data (uncleaned data) and confidential data (cleaned data) containing sensitive information.} For instance, in the summer of 2023, America's Datahub Consortium announced many requests for solutions and several involved generating synthetic data for privacy protections.\footnote{``Opportunities,'' Accessed on July 15, 2023. https://www.americasdatahub.org/opportunities/} One such request for solution\footnote{``Creation of Synthetic Data for the Survey of Earned Doctorates and Development and Use of Verification Metrics,'' Request for Solicitation from the National Science Foundation National Center for Science and Engineering Statistics.} states that ``the anonymity of the original dataset is not compromised since synthetic records do not correspond with real ones.'' This is the appeal of synthetic data for privacy protection. It lies in the concept of generating pseudo or fake records that maintain statistical representation of the confidential data. In other words, data users gain access to data that have similar structure and features as the confidential data without accessing sensitive information.

Published synthetic data are one of two ways data users access confidential data. The first approach is direct access, but such access is challenging to obtain due to limited availability for external users, strict eligibility requirements, and background checks. For example, to gain access to some federal data in the United States, a data user must be a U.S. citizen, submit a research proposal, complete an extensive background check, and use a secure enclave or environment to analyze the confidential data. The second approach involves using publicly released record-level data (e.g., synthetic data) and statistics (e.g., summary tables), which are more accessible through online repositories or web-based interfaces.

Although the synthetic data method has existed for over thirty years, the increase in computational power and methodological advancement to generate synthetic data and the rise of disclosure threats makes synthetic data very appealing. In recent years, several programming packages have emerged, such as \textit{synthpop} \citep{nowok2016synthpop} and \textit{tidysynthses}\footnote{“The tidysynthesis R package.” Presentation given at rstudio::conf(2022), Washington, DC, July 25 – 28.} for \verb|R|. Start up companies also offer propriety software to generate synthetic data for specific data types (e.g., healthcare\footnote{``Syntegra,'' Accessed July 15, 2023. https://www.syntegra.io/}) or for structured\footnote{``Gretal,'' Accessed July 15, 2023. https://gretel.ai/} and unstructured data\footnote{``Datagen,'' Accessed July 15, 2023. https://datagen.tech/}. Several U.S. federal government agencies are exploring synthetic data as a means to protect their data products. For example, the U.S. Census Bureau is actively investigating the creation of a synthetic version of the American Community Survey\footnote{``American Community Survey Disclosure Avoidance,'' Accessed July 15, 2023. https://www.census.gov/programs-surveys/acs/methodology/disclosure-avoidance.html}.

With synthetic data generation surging in popularity within both research and practical applications, an increasing number of government and private entities are turning to this approach as a means to protect confidential information. However, newcomers to the field may encounter a daunting amount of information, where some challenges and limitations of synthetic data not readily disclosed. This review will address these concerns by offering readers a comprehensive introduction to synthetic data for privacy protection. We provide an overview of the synthetic data along with the technical details of how to generate and evaluate synthetic data for privacy protection. We also discuss practical applications, give concluding remarks, and thoughts for future work.

\textbf{A note on terminology:} The data privacy community involves a diverse set of people from different technical and non-technical backgrounds. We define these group as \cite{williams2023promise} has:

\begin{itemize}
\item
  \textbf{Data users and practitioners:} individuals who consume the
  data, such as analysts, researchers, planners, and decision-makers.
\item
  \textbf{Data privacy experts or researchers:} individuals who
  specialize in developing data privacy and confidentiality methods.
\item
  \textbf{Data curators, maintainers, or stewards:} individuals who own
  the data and are responsible for its safekeeping.
\item
  \textbf{Data intruders, attackers, or adversaries:} individuals who
  try to gather sensitive information from the confidential data.
\end{itemize}

Also, given the various groups, we make a distinction between data privacy and data confidentiality. Generally, data privacy and data confidentiality mean different things. Data confidentiality refers to how the data privacy community protects participants' information in the data, such as who should have access to the sensitive data under what restrictions. Data privacy refers to the amount of personal information individuals allow others to access about themselves. For this article, we will use data privacy given its popularity, but readers should know there is a difference.

\section{Overview of synthetic data}
\label{sec:background}

%\subsection{Partial synthesis and full synthesis}
%\label{sec:background:partialfull}
In this section, we provide a concise overview of synthetic data. Additionally, we review and discuss important aspects of synthetic data, including the technical details of their generation and evaluation as well as use cases in Section \ref{sec:technical} and Section \ref{sec:app}, respectively.

The synthetic data approach to privacy protection largely originated from multiple imputation for missing data \citep{Rubin1987book}. In missing data problems, Bayesian models can be designed and estimated on the available data according to different missing mechanisms. Some examples include missing completely at random, MCAR; missing at random, MAR; and missing not at random, MNAR. Researchers and practitioners can impute the missing values from the posterior distributions estimated from these Bayesian models. Multiple imputed datasets are created to account for the uncertainty from model estimation and data generation. In contrast, for data protection problems, privacy experts and researchers can design and estimate the Bayesian models on the confidential data and then they generate the synthetic values from the posterior distributions estimated from the Bayesian models. Variables of observations that require privacy protection can then be replaced with synthetic values instead of their confidential variables in the public release.

Similar to multiple imputation for missing data problems, privacy experts and researchers create multiple synthetic datasets to account for the uncertainty in the model estimation and data generation process. Different combining rules for obtaining valid inferences from multiple synthetic datasets need to be used, which we will review in Section \ref{sec:technical:utility} in detail. Moreover, synthesis models for synthetic data generation do not need to be Bayesian as we will discuss in Section \ref{sec:technical:generation}. It is worth mentioning that the majority of the synthetic data literature deals with confidential data free of missing values. A few works, such as \citet{KimReiterKarr2018JOAS, Jiang2021JASA, YuHeRaghu2022JSSAM}, propose different approaches to simultaneously dealing with missing data imputation and synthetic data generation. Finally, the choice of the number of synthetic datasets to generate should be made specific to the application. Common choices include three, five, ten, and twenty synthetic datasets. 

From the beginning, two flavors of synthetic data exist: partially synthetic data and fully synthetic data. The main distinction between the two can be thought of as whether or not we need to protect all variables of all observations in the confidential dataset. If not, the data curator can choose the partially synthetic data approach, otherwise, the fully synthetic data approach. In both cases, multiple synthetic datasets should be generated to account for variability in the model estimation and data generation process. Different combining rules are required for partially and fully synthetic data.

For partially synthetic data, a synthetic dataset contains the same set of observations as in the confidential dataset, i.e., the two datasets have the same number of observations and there exists a one-to-one correspondence between observations in the two datasets \citep{Little1993synthetic}. The data curator replaces the sensitive variables with synthetic values either for all observations or for only a subset of observations, depending on the protection goals of the data curator. The remaining variables are left unchanged, or un-synthesized. 

Generating fully synthetic data can be thought of as all variables of all observations in the confidential dataset being replaced with synthetic values. \citet{Rubin1993synthetic} originally proposes that the generation process of fully synthetic data consists of two steps. The data curator first creates a fully synthetic population from the confidential dataset (treated as a sample), where all variables contain values generated from the synthesis model that is built and estimated on the confidential dataset. Second, the data curator takes a sample from the generated synthetic population to be released as a fully synthetic dataset. In this approach, the resulting fully synthetic dataset could have more or less observations compared to the confidential dataset. 

Nowadays, a fully synthetic dataset is often regarded as synthesizing all variables in the confidential dataset. The generation process would be similar to how a partially synthetic dataset is created, except that all the variables are synthesized from the synthesis model, whereas in a partially synthetic dataset, not all variables would be synthesized. We highlight the distinction early on given many misunderstandings of fully synthetic data in research and in practice. Moreover, as we will present and discuss in Section \ref{sec:technical:utility} for utility evaluation of synthetic data, different combining rules have been developed and are needed for obtaining valid inference for partially synthetic data and for fully synthetic data. If a fully synthetic dataset is generated using the partial synthesis approach (i.e., no synthetic population is generated), then combining rules for obtaining valid inference for partially synthetic data should be applied.

\section{Technical details of generating and evaluating synthetic data}
\label{sec:technical}

With a general understanding of synthetic data, we review and discuss the technical details of the synthetic data approach. We start with generating synthetic data in Section \ref{sec:technical:generation}, followed by their utility evaluation and disclosure risk evaluation, in Section \ref{sec:technical:utility} and Section \ref{sec:technical:risks}, respectively. Section \ref{sec:technical:tradeoff} discusses the risk-utility trade-off of synthetic data as part of their evaluation to help make optimal release choices.

\subsection{Synthetic data generation}
\label{sec:technical:generation}

As mentioned in Section \ref{sec:background}, the synthetic data approach mostly stemmed from multiple imputation for missing data problems. Although not all synthesis models need to be Bayesian, Bayesian models have played a large role in developing synthesis models since the beginning. In light of this, we focus on reviewing Bayesian synthesis models versus non-Bayesian synthesis models for synthetic data generation. There are other ways to categorize synthesis models. For example, parametric versus non-parametric.

Without loss of generality and due to popularity in practice and research, we focus on describing synthesis models for the purpose of generating partially synthetic data and fully synthetic data using the partial synthesis approach. Privacy experts and researchers create and estimate synthesis models using the confidential dataset. The partially synthetic datasets generated from these models contain the same observations in the confidential dataset. In other words, there is a one-to-one correspondence between records in the confidential dataset and in each synthetic dataset. On the other hand, the fully synthetic datasets generated from the partial synthesis approach are typically considered containing records that do not correspond to the same observations in the confidential dataset. In this case, the partial synthesis approach simply replaces all variables with synthetic values. More details about generating fully synthetic data according to the original proposal in \citet{Rubin1993synthetic} can be found in \citet{RaghuReiterRubin2003JOS, ReiterRaghu2007, Drechsler2011book}.

\subsubsection{Bayesian synthesis models}
\label{sec:technical:generation:Bayesian}

For all Bayesian synthesis models, the general approach to generating synthetic data lies in the use of the posterior predictive distribution. Essentially, we can think of synthetic data as predictions generated from the posterior distribution of the model parameters estimated on the confidential dataset. 

Let $\bm{Y}$ represent the confidential dataset containing $n$ observations and $p$ variables and $\bm{\theta}$ represent the parameters of the selected Bayesian model for $\bm{Y}$. Denote $f(\bm{Y} \mid \bm{\theta})$ as the sampling model and $\pi(\bm{\theta})$ as the prior choice for $\bm{\theta}$. By Bayes' theorem, the posterior distribution of $\bm{\theta}$ can be estimated by
\begin{equation}
    \pi(\bm{\theta} \mid \bm{Y}) \propto \pi(\bm{\theta}) f(\bm{Y} \mid \bm{\theta}). \label{eq:posterior}
\end{equation}

Through Markov chain Monte Carlo (MCMC) techniques of posterior estimation, we can generate posterior draws of parameters $\bm{\theta}$, denoted as $\tilde{\bm{\theta}}$. We can then simulate predictions of $\bm{Y}$ from the sampling model using the collection of posterior samples, $\tilde{\bm{\theta}}$. A synthetic dataset, denoted as $\tilde{\bm{Y}}$ sharing the same dimension as the confidential dataset $\bm{Y}$, can contain synthesized values of a subset of variables, and therefore being partially synthetic and correspond to the same observations as in ${\bm{Y}}$; or of all variables, and therefore being fully synthetic (i.e., using the partial synthesis approach) and do not correspond to the same observations as in ${\bm{Y}}$. 

When we need to generate multiple synthetic datasets, we first simulate $m$ independent sets of posterior parameters, $\tilde{\bm{\theta}} = \{\tilde{\bm{\theta}}^{(1)}, \cdots, \tilde{\bm{\theta}}^{(m)}\}$, from the posterior estimation process through MCMC. Next, each synthetic dataset, $\tilde{\bm{Y}}^{(l)}$ ($l \in 1, \cdots, m$), is generated from the sampling model using the set of parameter draws $\tilde{\bm{\theta}}^{(l)}$. We denote the resulting $m$ sets of synthetic datasets as $\tilde{\bm{Y}} = \{\tilde{\bm{Y}}^{(1)}, \cdots, \tilde{\bm{Y}}^{(m)}\}$. %We can see the clear resemblance of this process to multiple imputation for missing data problems using Bayesian models.

We can leverage the effective estimation of posterior predictive distributions of Bayesian models to generate synthetic data that resemble the features of the confidential data. The simplest examples would be using linear regression models to synthesize a continuous variable and logistic regression models to synthesize a binary variable.

In \citet{RosOlssonHu2020PSD}, the income variable is to be synthesized in a sample from the National Health Interview Survey. The goal is to create partially synthetic data with only income synthesized and every other variable remains un-synthesized. \citet{RosOlssonHu2020PSD} synthesizes the highly-skewed income variable using a two-phase synthesis approach by considering two forms. In the first phase, they model and synthesize a binary form using Bayesian logistic regression. In the second phase, the non-negative continuous is modeled and synthesized through Bayesian linear regression. In both phases, other variables in the sample are used as predictors and the aforementioned use of posterior predictive distributions is carried out for generating multiple partially synthetic datasets.

Apart from a straightforward illustration of using common Bayesian models for generating synthetic data, \citet{RosOlssonHu2020PSD} highlights the importance of careful design of Bayesian models by taking into account of the features of the confidential data. To demonstrate the superiority of the two-phase synthesis, they implement and compare the levels of utility preservation of a single-phase synthesis using Bayesian linear regression against the two-phase method. The two-phase synthetic data have a much higher resemblance of the confidential income variable. 

Recent literature highlights the potential of Bayesian modeling in synthetic data generation. With careful design and estimation of the Bayesian models, the resulting synthetic data can usually maintain high levels of utility. In what follows, we include several successful examples utilizing sophisticated Bayesian models.% compared to Bayesian regressions.

In the past decade, non-parametric Bayesian models have been extensively studied to protect samples consisting of multivariate un-ordered categorical variables. This line of work focuses on modeling the joint distribution of multivariate categorical variables using products of multinomial distributions through latent class modeling. Dirichlet process priors are used to exploit the appealing features of potentially infinite number of latent classes and therefore the ability of modeling any distribution \citep{DunsonXing2009JASA}. The term Dirichlet process mixtures of products of multinomials (DPMPM) has been used for this general synthesis strategy for multivariate un-ordered categorical data. \citet{HuReiterWang2014PSD} first applies the DPMPM synthesis model to a 2012 American Community Survey sample in the United States. \citet{drechsler2021synthesizing} utilizes the DPMPM synthesis model for a German large-scale administrative data where geographical information is to be protected. Similarly,  \citet{HuSavitsky2023TDP} uses the DPMPM synthesis model (and a non-parametric Bayesian areal level synthesis model that also relies on Dirichlet process priors) for synthesizing county labels to protect the county label information of consumer units in a 2017 Consumer Expenditure Surveys sample in the United States. \citet{CaoHu2022PSD} synthesizes information about youth risk behavior in a sample of the Youth Risk Behavior Survey in the United States using the DPMPM synthesis model.

In order to use the DPMPM synthesis model in practice, we need to be either working with a dataset consisting of un-ordered categorical only, or ready to discretize any non-categorical variables. Computation could be a concern compared to some non-Bayesian synthesis models, which we will discuss in Section \ref{sec:technical:generation:nonBayesian}. However, the \texttt{NPBayesImputeCat} R package \citep{NPBayesImputeCat} has made its usage much more possible with large datasets. Details of the synthesis procedure with illustrative synthesis examples using the \texttt{NPBayesImputeCat} R package can be found in \citet{NPBayesImputeCat2021RJournal}. Other challenges related to more complicated data features have been partially addressed in several extensions. 

One extension of the DPMPM synthesis model is \citet{ManriqueHu2018JRSSA}, which deals with structural zeros (i.e., impossible combinations of variables, such as a married 4-year-old) within the DPMPM. %Another is \citet{HuHoshino2018PSD}, where the authors replace the multinomials with Quasi-Multinomials in the DPMPM to allow a tuning parameter for balancing utility-risk trade-off. 
To protect household data with nested individuals within households, \citet{HuReiterWang2018BA} designs a nested version of the DPMPM, a two-level latent class model, that accounts for the dependencies among household members. The \texttt{NestedCategBayesImpute} R package \citep{NestedCategBayesImpute} can be used for the nested DPMPM synthesis models, which also accommodates structural zeros in nested data. We refer interested readers with a background in multiple imputation to \citet{AkandeReiterLi2017}, which compares the DPMPM as an imputation engine to other imputation models based on chained equations.% for un-ordered categorical data. 

Synthesizing geographical information using Bayesian spatial models has been another well-studied topic in recent literature. This line of work focuses on utilizing features of areal level spatial models, also known as disease mapping models \citep{ClaytonKaldor1987}. It relies on dividing the area of interest into a grid consisting of multiple cells and creating multiple distinct attribute patterns through cross-tabulation of observations' non-spatial (and typically categorical) attributes. We can use Poisson regression models to estimate the number of observations in a given cell with a given attribute pattern. After MCMC estimation, we then make posterior predictions the numbers of observations, and synthetic geographical information, such as individual locations, can be generated according to the normalized probability vectors of the given attribute pattern within the grid cell.

\citet{Paiva2014SM} induces spatial correlation among neighboring grid cells with the intrinsic conditionally autoregressive priors for spatial random effects in Poisson regression and illustrates the method with an application to create partially synthetic locations for a subset of North Carolina mortality records from 2002. In the aforementioned \citet{HuSavitsky2023TDP}, a similar Bayesian areal level spatial model is considered. However, due to little geographic information carried in county labels in the Consumer Expenditure Surveys sample, the authors include non-spatial random effects and use Dirichlet process priors. \citet{QuickHolanWikle2018JRSSA} proposes differential smoothing techniques to generate partially synthetic geocoded data.

These applications show that spatial models are not necessarily used or preferred when synthesizing geographical information (another example is the aforementioned \citet{drechsler2021synthesizing}). The model choice should depend on specific features of the data to be protected and computation consideration, among other things. Despite their usefulness, areal level spatial synthesis models face a significant challenge in practice in their need of computation resources, especially when having a not-too-large number of distinct attribute patterns. This issue is also seen for the use of Bayesian marked point process models in \citet{QuickHolanWikleReiter2015SS}, where the model estimation can be computationally intractable for an application of the same and relatively small 2002 North Carolina mortality rates sample as in \citet{Paiva2014SM}. This leads the authors to suggest several simplifications of the model. 

There has also been work using Bayesian networks as synthetic data generation models. \citet{Young2009JOS} proposes the general framework of utilizing Bayesian networks and \citet{Kaur2021JAMIA} includes examples of using Bayesian networks to synthesize health data. Existing tools, such as the \texttt{bnlearn} R package \citep{bnlearn} applied by \citet{Kaur2021JAMIA}, can be used to perform Bayesian networks estimation from which synthetic data can be created.

When more than one variable is to be synthesized using univariate Bayesian models (i.e., one synthesis model for one variable), we can use the sequential synthesis approach, which will be reviewed in detail in Section \ref{sec:technical:generation:nonBayesian}.

Since synthetic data generation essentially makes predictions from the estimated posterior distributions, we can either write our own MCMC algorithms or leverage existing tools, such as Stan \citep{Stan} and JAGS \citep{jags}, for MCMC estimation. These routes are necessary for advanced Bayesian models, unless there exists tools for their MCMC estimation. As an example, the \texttt{NPBayesImputeCat} R package is for estimating the aforementioned DPMPM model. For more widely-used Bayesian models, such as linear regressions and certain GLMs, R packages including \texttt{brms} \citep{R-brms} and \texttt{rstanarm} \citep{R-rstanarm} can be used for MCMC estimation. In all these approaches, we can extract posterior parameter draws from the model fits and generate synthetic data with just a few lines of code or a function to streamline the process.

\subsubsection{Non-Bayesian synthesis models}
\label{sec:technical:generation:nonBayesian}
We can also generate synthetic data from other types of models that are not in a Bayesian framework. We roughly group these methods as nonparametric (e.g., synthetic data generation based from an empirical distribution) or parametric (e.g., synthetic data generation based from a parametric distribution or generative model).

Similar to Bayesian synthesis models, when implementing a parametric approach, it becomes crucial to select a suitable model that accurately represents the confidential data to preserve as many of the underlying data relationships as possible. One straightforward parametric method for generating synthetic data is selecting an appropriate probability distribution. This approach involves making random draws from the chosen distribution based on the parameters or sufficient statistics from the confidential data. For instance, if the confidential data follows a Gaussian distribution, we can generate the synthetic data by drawing random samples from a normal distribution, utilizing the mean and variance obtained from the confidential data.

More complex models may include using prediction models, such as regression, or conducting a sequential synthesis that estimates models for each predictor with previously synthesized variables used as predictors. The latter approach captures more of the multivariate relationships (or joint distributions) without being too computationally expensive, as compared to estimating a complicated joint distribution of all predictors to be synthesized. We can select the synthesis order based on the priority of the variables or the relationships between them. Typically, the earlier in the order a variable is synthesized, the better the confidential information is preserved in the synthetic data. \citet{bowen2021differentially} proposes a method that ranks variable importance by either practical or statistical utility and sequentially synthesizes the data accordingly.

For the nonparametric approach, the most basic method is using marginal tables and random sampling. For instance, we can categorize the data into groups or bins, calculate the proportion, and randomly sample synthetic values based on the proportion. This approach is simple and quick to implement, but does not properly capture the variability and relationships of continuous variables because the continuous values get discretized \citep{BowenLiu2020SS}.

A more complex nonparametric technique is a sequence of Classification and Regression Tree (CART) models \citep{gordon1984classification}. \cite{reiter2005using} originally proposes to use a collection of nonparametric models to generate partially synthetic data. Essentially, CART iteratively divides the data using binary splits until reaching homogeneous nodes. If the target variable is categorical, CART employs classification trees to predict the outcome that constructs the tree by progressively partitioning the data into binary segments. For continuous variables, CART uses regression trees to determine the splitting value that separates the continuous values into partitions. A regression tree generates nodes with the lowest sum of squared errors, calculated as squared deviations from the mean.

Mathematically, we define the sequence of CART models as follows:
\begin{align}\label{eq:seq}
    \begin{split}
        f(\boldsymbol{Y}\mid\boldsymbol{\theta})&=f(Y_1,Y_2,...,Y_p\mid\theta_1,\theta_2,...,\theta_p,\boldsymbol{X}) \\
        &= f_1 (Y_1\mid\theta_1,\boldsymbol{X})\cdot f_2(Y_2 \mid Y_1,\theta_2,\boldsymbol{X})...f_p(Y_p \mid Y_1,...,Y_{p-1},\theta_p,\boldsymbol{X}) 
    \end{split}
\end{align}
\noindent where $Y_k$ for all $k = 1,...,p$ are the variables to be synthesized, $\boldsymbol{\theta}$ are vectors of model parameters, such as regression coefficients and standard errors, $p$ is the total number of variables, and $\boldsymbol{X}$ represents the un-synthesized predictor variables.

Overall, CART tends to offer greater flexibility compared to parametric approaches like regression-based models, allowing it to account for atypical variable distributions and nonlinear relationships that may be challenging to explicitly identify and model. Recent studies have demonstrated its superior performance over parametric regression-based methods while remaining computationally feasible \citep{bowen2022synthetica, bonnery2019promise, drechsler2021synthesizing}.

\subsection{Evaluation of data utility}
\label{sec:technical:utility}
As one might imagine, synthetic data must be sufficiently useful for public release in lieu of confidential data. Privacy experts and researchers refer to this process as evaluating the data utility, where higher utility typically corresponds to greater accuracy and usefulness of the data. This process is a critical factor in selecting the most suitable synthetic data generation method.

If we release multiple synthetic datasets, data users will need use combining rules to obtain the proper statistical inference. Suppose the parameter of interest is $\beta$. Denote the estimate of $\beta$ in the $l{th}$ synthetic data by $\hat{\beta}_l$ and the associated standard error by $v_l$. The final point estimate $\hat{\beta}$, for $\beta$, is
\begin{equation}
    \bar{\beta} = \textstyle{m^{-1}\sum_{l=1}^m}\hat{\beta}_l
\end{equation}
with Var$(\bar{\beta})$ estimated by
\begin{equation}
    T_p = m^{-1}B + W,
\end{equation}
for partially synthetic data (or fully synthetic data generated from the partial synthesis approach), where $B= \sum_{l=1}^m (\hat{\beta}_l-\bar{\beta})^2 /(m-1)$ is the between-set variability and  $W=m^{-1}\sum_{l=1}^m v^2_l$ is the average per-set variability. Data users can also use a $t$ distribution with degrees of freedom of $v_p = (m-1)(1+W/(B/m))^2$ to obtain confidence intervals and conduct hypothesis tests, among other things.

For fully synthetic data generated, \citet{Rubin1993synthetic} originally proposes we use $T_f = (1 + 1/m)B - W$ for the variance estimate of $\hat{\beta}$ (see \citep{Reiter2002} for an alternative non-negative variance estimator) and $v_f = (m - 1)(1 - W/((1 + 1/m)B))^2$ for the degrees of freedom of corresponding the $t$ distribution. We refer interested readers to \citet{RaghuReiterRubin2003JOS, ReiterRaghu2007, Drechsler2011book} for further details on these combining rules. Additionally, \citet{raab2017practical} provides an overview of other ways to estimate the variance and recommendations on which to use under different settings. The reader should note that the combining rules become more complex for other types of inferences, such as Chi-squared tests \citep{li1991significance}.

For simplicity, in the following we outline the various utility metrics of a single synthetic dataset.

\subsubsection{Global Utility}
One of the most common ways to evaluate the synthetic data is assessing the univariate and multivariate distributional similarities between the confidential data and the synthetic data. Privacy experts and researchers refer to these measures as global utility (also known as general utility).

For the univarate case, we could calculate the frequencies and relative frequencies of the categorical variables in the synthetic and confidential data. When the variables are numeric, we could compute the means, standard deviations, skewness, kurtosis, percentiles, and number of zero/non-zero values. We can also visually compare the the results of the univariate distributions from the synthetic and confidential data using a histogram, density plots, or empirical cumulative distribution function (eCDF) plots. \citet{Woo2009JPC} proposes the use of the eCDF as a global utility metric, which is particularly useful for univariate numeric variables. While the metric can be easily extended to multivariate distributions, implementing the eCDF estimation itself in multivariate cases is not straightforward. Therefore, the eCDF global utility metric is rarely used beyond univariate cases. 

The eCDF approach that \citet{Woo2009JPC} proposes belongs to a group of methods called discriminant based methods that measure how well a predictive model can distinguish or discriminate between the records from the confidential and synthetic data, assessing the relationships among multiple variables. In other words, these methods combine the synthetic and confidential datasets and see how well a predictive model can discriminate between synthetic observations and confidential observations in the combined dataset. The inability to distinguish between the records suggests a high utility synthesis, because the synthetic observations closely resemble the confidential observations. Any predictive model can be used, such as logistic regression and CART.

Most discriminant based methods are propensity score based, allowing the method to compare the similarity of two datasets of the same structure of any dimension without making assumptions on the distributions of the attributes. Mathematically, these methods use the following steps. Let $\bm{Y}$ be the confidential dataset with $n$ observations and $p$ variables.
\begin{enumerate}
    \item Combine the confidential and synthetic datasets, each of size $n$. Create an indicator variable $T$ where $T_i=1$ if  record $i$ is from the synthetic data and $T_i=0$ otherwise  for $i=1,\ldots, 2n$.
    \item Calculate the propensity score for each record $i$, $e_i=\Pr(T_i=1 \mid Y_i)$, through a classification algorithm, with the data attributes as input features.
\end{enumerate}
What is done with the propensity scores next depends on the discriminant based method. \citet{Woo2009JPC} computes the mean squared error (MSE) of the propensity score against the true proportion of synthetic cases. \citet{Snoke2018JRSSA} enhances \citet{Woo2009JPC}'s approach by computing the average MSE between the propensity scores and the expected probabilities called the propensity score mean squared error (pMSE). Essentially, pMSE normalizes the MSE statistic by its expected null value and standard deviation, helping with its interpretability and differentiating the synthetic dataset apart from the confidential dataset. \citet{Snoke2018JRSSA} also develops the pMSE ratio, which is one of the most popular discriminant based methods. The pMSE ratio is the average pMSE score across all records, divided by the null model, where the null model is the the expected value of the pMSE score under the best case scenario when the model used to generate the data reflects the confidential data perfectly. \citet{sakshaug2010synthetic} discretizes the propensity scores based on how the Chi-squared test is formulated. Finally, \citet{bowen2021differentially} calculates the eCDFs of the propensity scores of the synthetic and confidential data and then computes the the KS (Kolmogorov-Smirnov) distance, a method called SPECKS. In other words, the SPECKS method considers the worst-case separation between the synthetic dataset and the confidential dataset.

What the discriminant based metrics \textit{actually} measures for assessing the synthetic data quality varies depending on the method and the classification algorithm. For instance, \citet{bowen2021comparative} compares several utility metrics, such as the pMSE ratio and SPECKS, to evaluate differentially private synthetic datasets\footnote{Differentially private synthetic data are synthetic data that satisfies the definition of differential privacy, which quantifies the disclosure risk in formal ways. To learn more about differential privacy and formal privacy, we refer interested readers to \citet{williams2023promise}.} for a data challenge. The authors find that the utility metric algorithms produce mixed results in ranking the best performing differentially private synthetic data method. Conducting a study to analyze what features of the synthetic data are captured by various discriminant based methods using different classification models would be invaluable to the field \citep{drechsler2022challenges}. However, to the best of our knowledge, no such study exists for synthetic data with and without differential privacy or formal privacy guarantee.

\subsubsection{Analysis-specific utility}
Analysis-specific utility measures the similarity of results between confidential and synthetic datasets for a specific analysis or multiple analyses. Simply put, these metrics assess if data users would reach the same conclusions whether applied to the confidential dataset or synthetic dataset. The specific utility metrics will vary across applications, depending on the common uses of the data.

There are a few ways to compare the synthetic and confidential data outputs. If the analysis involves totals or proportions, \citet{taub2020impact} develops the ratio of estimates (ROE). ROE is the ratio of the confidential and synthetic data estimates such that the smaller of the two estimates is divided by the larger one. If the ROE value is 1, then the confidential and synthetic estimates are the same. For univariate estimands, such as regression coefficients and means, \citet{KarrKohnenOganianReiterSanil2006} creates the confidence interval overlap (CIO). This metric is commonly seen in the synthetic data literature, which compares the confidence intervals (CIs) from the confidential and synthetic datasets to see how much the synthetic data generation affects inference. \cite{Snoke2018JRSSA} propose a modification that allows for negative CIO values that show how far off the confidence intervals do not overlap. We define the measure as:

\begin{equation}\label{eqn:io}
CIO = 0.5 \bigg( \frac{min(u_c, u_s) - max(l_c, l_s)}{u_c - l_c} + \frac{min(u_c, u_s) - max(l_c, l_s)}{u_s - l_s} \bigg)
\end{equation}

\noindent where $u_c$, $l_c$ and $u_s$, $l_s$ are the upper and lower bounds for the confidential and synthetic CIs respectively. The metric measures how much the CIs estimated the confidential and synthetic data overlap for a single estimate on average, where the maximum value is 1. The value is negative if the intervals do not overlap and grows more negative as they move further away from each other.

A drawback to the CIO measure is the inability to distinguish whether the confidential or the synthetic dataset has a wider CI that covers the other interval. If one interval is wider but completely encompasses the other interval, the minimum value is 0.5 regardless of the width. This is why \cite{barrientos2021feasibility} creates a new metric called sign, significance, and overlap (SSO) match. SSO is the proportion of times that intervals overlap and have the same sign and significance. Although created for evaluating differentially private regression outputs, the SSO can be applied to synthetic data outputs as well.

\subsubsection{Fit-for-Purpose:}
The final group of utility metrics are called fit-for-purpose and are not discussed as often in the literature. \citet{drechsler2022challenges} states how fit-for-purpose measures could be considered something in between the previous two utility metric types. In other words, fit-for-purpose metrics are not global measures, because they focus on certain features of the data, but may not be specific to an analysis that data users are stakeholders are interested in like analysis-specific utility metrics. 

Privacy experts and researchers use fit-for-purpose metrics to address the limitations of other metric types. \citet{drechsler2022challenges} shows how a global utility metric (standardized pMSE, another variation of the pMSE metric) is dependent on the classification model and misses key differences in the synthetic and confidential datasets. This work highlights how global utility metrics can be too broad and miss aspects of the synthetic dataset that do not align with the confidential dataset. On the other hand, analysis-specific metrics may perform well for the selected analyses on the synthetic data but not for others. This is why it is critical to determine the proper analysis, but it is difficult to anticipate all downstream data uses. For example, decennial data census products in the United States are utilized in thousands of different ways, making it impossible to predict all potential use cases.

Therefore, fit-for-purpose metrics help privacy experts and researchers assess if their synthesis makes sense before implementing the other utility metrics. Some examples include ensuring population totals or ages are positive. The discriminant based metrics could be used as a fit-for-purpose metric if only a subset of the variables are included in the classification model for the propensity score matching \citep{raab2017guidelines}.

\subsection{Evaluation of disclosure risks}
\label{sec:technical:risks}

Synthetic data should be considered for public release based on the amount of disclosure risks they carry in addition to the level of utility preservation. In other words, synthetic data with a high level of utility and also a high level of disclosure risks might not be an ideal candidate for public release. It is therefore paramount to define disclosure risks, and, subsequently, design metrics and develop evaluation methods that are relatively easy to implement computationally. Data curators should evaluate and compare the disclosure risks of the confidential data and of the synthetic data. This process demonstrates the amount of reduction of disclosure risks the synthetic data can offer and can serve as a criterion for public release. As such, disclosure risks can be defined and evaluated on the microdata, including the confidential and the synthetic.

We focus on two types of disclosure risks that are commonly considered for synthetic data research and applications. The first type is called identity disclosure, which is sometimes known as identification disclosure or re-identification disclosure. Identity disclosure risk considers the scenario where a malicious intruder attempts to identify record(s) of interest in the confidential data. The second type is called attribute disclosure, which refers to the scenario where an intruder seeks to infer the confidential values of variable(s) of interest for targeted record(s). 

In both cases, the data intruder would use available information or knowledge from other sources, combined with the published microdata (the confidential or the synthetic, depending on which are being evaluated) to figure out the identity or the confidential value(s) of the targeted record(s). Often times, an identity disclosure could lead to an attribute disclosure, whereas an attribute disclosure can happen without an identify disclosure in the first place. \citet{Hu2019TDP} provides a review of Bayesian estimation of these two types of disclosure risks for synthetic data. We will include additional non-Bayesian evaluation methods in the upcoming sections.

Readers should be aware that the literature will sometimes describe a third type of disclosure risk known as ``inferential disclosure risk'' \citep{shlomo2018statistical, matthews2011data, fcsm2005}. However, our review specifically concentrates on identity and attribute disclosure risks. It is important to note that how we define attribute disclosure risk encompasses certain aspects that are referred to in the literature as inferential disclosure risk.

When generating and evaluating multiple synthetic datasets, it is common to perform risk evaluation on each individual synthetic dataset. Afterwards, we combine the risk results across the datasets by taking the averages to report the disclosure risks of multiple generated synthetic datasets. In our forthcoming review, for simplicity, we focus on describing the disclosure risk evaluation of a single synthetic dataset.

\subsubsection{Identity disclosure risk}

As previously mentioned, identity disclosure happens when an intruder chooses to use certain available information or knowledge from other sources and the published microdata to attempt identification of target record(s) in the confidential data. Identity disclosure is only considered for partially synthetic data, because the synthetic records correspond to the same set of records in the confidential data, resulting in a one-to-one mapping between the datasets \citep{Hu2019TDP}. In fully synthetic data, no records correspond to any records in the confidential data, and therefore an identify disclosure will not be possible and is not considered.

\citet{ReiterMitra2009} proposes a general framework to evaluate identity disclosure risks in partially synthetic data based on Bayesian probabilistic matching. In this framework, we assume an intruder will attempt to identify a target record in the confidential dataset with available information from other sources. %Such information could be the values of a subset of the target record's variables through their personal knowledge or from other data sources. 
Within this subset of variables, some variables might be synthesized in the synthetic dataset. This means those values will be different from those in the confidential dataset, while others stay un-synthesized. Even for the un-synthesized values, the intruder's information about the target record might not be accurate. Considering the intruder's behavior becomes vitally important, such as what information they have access to and how they would use it to identify the target record. We can easily see this framework assumes the knowledge and behavior of the data intruder when attempting identification of target record(s). As a result, identity disclosure risk results of the same synthetic data can vary depending on what assumptions are made. 

We can evaluate identity disclosure risks of the synthetic dataset using the framework in \citet{ReiterMitra2009} in its simplest form. This involves directly matching a selected set of variables an intruder would use between records in the synthetic dataset with a target record in the confidential dataset. For simplicity, we first consider the selected set of variables being categorical. Suppose using the selected set of categorical variables, where there are $c$ records in the synthetic dataset being matched exactly with the target confidential record. Further, suppose that the true record (i.e., the synthetic record sharing the same identity as the confidential target record) is among the $c$ matched records. At this point, the intruder would randomly select one of the $c$ matched records as the identity of the target record, which produces a probability of $1/c$, a measure of the identity disclosure risk of this target record. However, if the true record is not among the $c$ matched records, then there will be a $0/c = 0$ identity disclosure probability for this target record. An identity disclosure probability of 0 could easily happen if the intruder utilizes a variable that is synthesized to a different value or they simply have inaccurate information about an un-synthesized variable. At the file-level, we can take the sum of all individual-level identity disclosure probabilities and present a summary statistic. This is referred to as the expected match risk.

In this framework, there are two other commonly used file-level identity disclosure risk summaries: the true match rate and the false match rate. The true match rate is the percentage of true unique matches (i.e., only one matched record $c = 1$ and it is the true record) among all target records. The false match rate is the percentage of false unique matches (i.e., only one matched record $c = 1$ and it is not the true record) among all unique matches. A higher expected match risk is associated with a higher disclosure risk. Higher true match rates also entail higher disclosure risk. Similarly, higher false match rates lead to higher disclosure risk.
We can also evaluate the identity disclosure risks of the confidential dataset in a similar fashion and subsequently compare the results between the confidential dataset and the synthetic dataset for a relative evaluation.

A variation of \citet{ReiterMitra2009}'s framework is \citet{HornbyHu2021TDP}, which evaluates the usage of a radius concept when matching for a continuous variable. Instead of exact matching for a categorical variable, we will declare a match when a continuous variable of the synthetic value falls within an interval with a certain width based on the chosen radius of its confidential value. \citet{HornbyHu2021TDP} further investigates the effects of the choice of the radius on the identity disclosure risks, among other things, and make the general recommendation of using a percentage radius instead of a fixed-value radius so that the matching can be made proportional to the confidential value. %Similar concepts are explored in \citet{HuSavitskyWilliams2022rebds} when tuning the utility-risk trade-off of synthetic data. 
The R package \texttt{IdentificationRiskCalculation} \citep{IdentificationRiskCalculation} can be used for identity disclosure risk evaluation using the simplest form of the \citet{ReiterMitra2009} framework for various data types.

Record linkage methods, developed mainly for the purpose of linking records from multiple datasets, can also be used for identity disclosure risk evaluation \citep{Winkler2004PSD, TorraAbowdDomingo2006PSD}. For this approach, we assume the data intruder seeks to link records between the confidential and the synthetic datasets. Among the established links between records from the two datasets, we can calculate the percentage of true links as an identity disclosure risk metric. True links are cases where the two records (i.e., one in the confidential dataset and the other in the synthetic dataset) refer to the same individual are correctly linked.

Similar to the framework proposed by \citet{ReiterMitra2009}, record linkage methods also use utilize variables, called keys in the record linkage literature, when performing the linkage. The similarity extends to the next stages of record linkage. As an example, what information and behavior the intruder would use to compare generated pairs based on the selected keys. In other words, the record linkage approach to identity disclosure risk evaluation also depends on assumptions of intruder's knowledge and behavior. We can use various existing record linkage algorithms in the evaluation process. For example, the \texttt{reclin} R package \citep{reclin} implements the expectation-maximization algorithm for probabilistic record linkage \citep{FellegiSunter1969JASA} and allows one-to-one linkage between datasets. In one-to-one linkage, one record in the confidential dataset is linked to one and only one record in the synthetic dataset, which could be desirable in identity disclosure risk evaluation. 

\subsubsection{Attribute disclosure risk}

Attribute disclosure occurs when an intruder chooses to use certain available information or knowledge from other sources, along with the published microdata, to infer the confidential value of target record(s) in the confidential data. Recall that an identity disclosure could lead to an attribute disclosure, but an attribute disclosure can happen without an identity disclosure. Therefore, privacy researchers considers attribute disclosure for both partially synthetic and fully synthetic data. 

The majority of the literature focuses on the evaluation of the attribute disclosure risk without a prior identity disclosure for partially synthetic data. One main line of work uses statistical models based on available variable(s) to predict the value(s) of other variable(s) to infer the attribute(s) of interest. We will call the available variable(s) used in this process as the key variables, while the other variable(s) of attribute interest as the target variable(s). The main difference between different approaches in this area of work is what types of statistical models are used in the prediction task. Without loss of generality, we next describe the approaches for predicting one categorical target variable with categorical key variables. The extension to continuous key variables could be achieved through a similar approach, starting from exact matching and moving to a radius-based matching, as in \citet{HornbyHu2021TDP} for identity disclosure risk evaluation.

The crudest approach is to use the empirical distribution of the target variable to predict the value for a confidential record of interest. We first match records from the synthetic dataset, which share the same categorical key variable(s) as the confidential record of interest. Among these matched synthetic records, we next calculate percentage of synthetic records who share the same value of the target variable as the confidential record. The resulting percentage can be thought of a probability of attribute disclosure, which is called the individual correct attribution probability (CAP) \citep{Elliot2014CMIST, Taub2018PSD, BaillargeonCharest2020PSD}. We can perform the evaluation process for each confidential record of interest and take the sum and the average as the file-level CAP summaries. While easy to implement in most settings, the empirical distribution based CAP statistics are not leveraging more advanced statistical modeling techniques, which can potentially help the intruder for better predictions. Consequently, the CAP statistics might be overly conservative and underestimate the attribute disclosure risks given their simple prediction process.

Due to its prediction nature, privacy experts and researchers have since proposed to implement advanced statistical models, including machine learning techniques, as a means to evaluate attribute disclosure risks based on prediction. In fact, many classification algorithms can be easily turned into attribute disclosure risk evaluation methods. These approaches would also involve the use of key variables as predictors and thus make assumptions of intruder's knowledge and behavior. \citet{Choi2017PMLR} and \citet{Kaur2021JAMIA} utilize the $k-$nearest neighbor classification algorithm for predicting the values of the target variable for a confidential record of interest. Similar to the CAP statistics, we can then calculate the percentage of records for which the prediction is correct. This measure can be used as an attribute disclosure risk metric. 

Privacy experts and researchers have also dealt with evaluating attribute disclosure risk without a prior identity disclosure for fully synthetic data. One line of work considers the worst case scenario, where the intruder knows about every other record in the confidential dataset except for the record of interest, and then attempts to infer the confidential values of its synthesized variables \citep{Reiter2014framework, HuReiterWang2014PSD}. We can consider such a worst-case-scenario approach providing the most conservative evaluation of the attribute disclosure risk. In other words, any less information about the rest of the confidential dataset, the evaluation would result in a lower level of attribute disclosure risk. However, this worst-case-scenario approach has limitations to its sophisticated implementation procedure, especially when less information is assumed to be known, making it less computationally feasible. Typically, this approach requires either re-estimation of the posterior distribution without the last and unavailable record of interest or some approximation solution that avoids rounds of re-estimation. We refer interested readers to \citet{HuReiterWang2014PSD} for an application for evaluating attribute disclosure risk of multivariate fully synthetic categorical data and \citet{WeiReiter2016} for an application with multivariate fully synthetic continuous data. \citet{WangReiter2012} and \citet{Paiva2014SM} include applications of a similar approach for partially synthetic data.

Lastly, it is important to highlight the flexibility of defining attribute disclosure risk metrics that are tailored to the features of the confidential data that need to be protected. For example, when working with geographies, \citet{WangReiter2012} develops a couple of geographies-specific attribute disclosure risk metrics, including a Euclidean distance between the intruder's inferred value of the longitude and latitude and the confidential longitude and latitude. Similarly, \citet{Paiva2014SM} proposes a file-level risk metric of the records with the true location being the maximum posterior probability of the confidential record, among other things. For establishment data, \citet{Kim2015JOS, WeiReiter2016, KimReiterKarr2018JOAS} consider a scenario where the intruder with access to the information about the second largest value of a certain variable (e.g., payroll) attempts to use the synthetic data to learn about the establishment with the largest value of the same variable. \citet{Mitra2020TDP} deals with attribute disclosure risks that exist in longitudinal data. Considering the diversity of contexts and successes demonstrated in these works, it is prudent for privacy experts and data curators to design their attribute disclosure risk metrics and evaluation techniques based on the specific data context.

\subsection{Risk-utility trade-off}
\label{sec:technical:tradeoff}

Synthetic data resembling the confidential data would have higher utility, while at the same time, higher disclosure risks, whereas synthetic data with less resemblance would result in lower utility and lower disclosure risks. In fact, this risk-utility trade-off exists not only in synthetic data, but also in a variety of other data protection techniques. The earliest work demonstrates such a trade-off appears in \citet{DuncanStokes2004CHANCE}, where the top-coding protection technique is investigated in terms of utility preservation and disclosure risk reduction and the balance of their trade-off. Once the R-U (risk-utility) confidential map concept is proposed, privacy experts and data curators have since picked it up in the synthetic data literature. We can use a R-U confidential map to determine which one (or several ones) exhibit the most optimal risk-utility trade-off when it comes to determining the public release.

In practice, if we are tasked with protecting a given confidential dataset, we can propose a few synthetic data generation methods and perform utility and disclosure risk evaluation. We can then pick the one with the most optimal balance for public release. Some research attempt to tune the risk-utility trade-off directly in the synthesis model. This approach eliminates the need to develop one or more brand new synthesis models when the existing synthesis model produces synthetic data with high utility but also high disclosure risks. We can instead tune this existing synthesis model to a version that produces synthetic data with a lower level of disclosure risk at the price of a reduced level of utility. 

\citet{HuHoshino2018PSD} uses the Quasi-Multinomial distribution, which includes an additional parameter compared to the multinomial distribution. This parameter can effectively tune the risk-utility trade-off of the resulting synthetic multivariate categorical data. \citet{Jiang2021JASA} incorporates a tuning parameter in their mask component that can balance the amount of utility preservation and disclosure risk of the resulting synthetic data. \citet{Jackson2022JRSSA} also integrates tuning parameters in their saturated models to generate categorical synthetic data that can balance between utility and disclosure risks. Extensions to their work can be found in \citet{jackson2022integrating}. Another example is \citet{HuSavitskyWilliams2022rebds}, where a record-index risk probability (between 0 and 1) is calculated based on the confidential data. This probability is then subsequently used in the likelihood expression through the exponents, such that high-risk records will be downweighted in posterior estimation. This results in the records receiving a higher level of protection in the resulting synthetic data. Finally, \citet{SchneiderHuMankadBale2023ESWA} proposes a Bayesian generalized linear model synthesis model with a shrinkage prior that allows the tuning of risk-utility trade-off for user-generated content data. The CART synthesis models are also investigated in \citet{SchneiderHuMankadBale2023ESWA}, where the complexity parameter of the CART models is experimented to tune risk-utility trade-off of the resulting synthetic data.

\section{Applications and Use Cases}
\label{sec:app}

In this section, we first provide lists of synthetic data applications and use cases classified by the characteristics of the confidential data to be protected in Section \ref{sec:app:summary}. In each, we include references and encourage the readers to refer to those works for further details. We then discuss two examples in detail in Section \ref{sec:app:examples}.

\subsection{Summary}
\label{sec:app:summary}
Most notably, the U.S. Census Bureaus implements and publishes several synthetic data products for public use. These include OnTheMap \citep{OnTheMap2008}, the Synthetic Longitudinal Business Databases \citep{SynLBD2011, SynLBD2014}, and the Survey of Income and Program Participation synthetic beta files \citep{SIPPdocument}.

When dealing with multivariate un-ordered categorical data synthesis, there are the DPMPM model \citep{HuReiterWang2014PSD, HuHoshino2018PSD, CaoHu2022PSD} with extensions to household data \citep{HuReiterWang2018BA} and structural zeros \citep{ManriqueHu2018JRSSA}. To synthesize geographical information, we can either use spatial models \citep{Paiva2014SM, QuickHolanWikleReiter2015SS, QuickHolanWikle2018JRSSA} or non-spatial models \citep{WangReiter2012, drechsler2021synthesizing, HuSavitsky2023TDP}. 

Examples specific to application fields include: Bayesian networks models have shown some potential \citet{Kaur2021JAMIA} for synthesizing health data; different synthesis models, Bayesian or non-Bayesian, have been proposed to synthesize establishment data \citep{DrechslerDundlerBenderRasslerZwick2008, DrechslerBenderRassler2008, DrechslerReiter2009JOS, WeiReiter2016, KimReiterKarr2018JOAS, ThompsonKim2022JSSAM} (some of these incorporates extensions, such as edit-imputation); methods to incorporate the sampling weights when creating synthetic survey data \citep{DrechslerReiter2010, HuSavitskyWilliams2021tabular, YuHeRaghu2022JSSAM}; for administrative data synthesis, there are \citet{drechsler2021synthesizing} and \citet{Jackson2022JRSSA}, which include both Bayesian and non-Bayesian options; finally, sequential CART models have synthesized complex tax \citep{bowen2022synthetica, bowen2022syntheticb, bowen2020synthetic} and education data \citep{goldstein2020expanding,bonnery2019promise}.

\subsection{Two detailed examples}
\label{sec:app:examples}

\subsubsection{Partially synthetic data with Bayesian models}
\label{sec:app:examples:partial}

Our first example is a partially synthetic data application for a sample of Airbnb listings in New York City with Bayesian models. \citet{GuoHu2023TAS} investigates the use of synthetic data to protect two sensitive variables in the sample. The first variable is the number of available days of an Airbnb listing, a count variable which has a large amount of zero-valued records. This variable is also truncated at the two ends at 0 and 365. The other variable is the price, which is numeric. Bayesian models are selected for a sequential synthesis of these two sensitive variables, with a zero-inflated truncated Poisson regression model for the number of available days first and a linear regression model for price next. Predictors including neighborhood (categorical), room type (categorical), and review count (count) are used in both models. MCMC estimation software JAGS is used in model estimation. The authors then use several utility metrics. For global utility, they use pMSE, eCDF, and another cluster analysis based metric \citep{Woo2009JPC}, whereas for analysis-specific utility, they evaluate the mean, extreme quantiles, and several regression coefficients with CIOs. In addition to evaluating both identity disclosure and attribute disclosure, the authors further experiment with uncertainties in intruder's knowledge and behavior in identity disclosure risk evaluation.

\subsubsection{Fully synthetic data with non-Bayesian models}
\label{sec:app:examples:full}
Our next example involves generating a fully synthetic dataset for U.S. taxpayer data using sequential CART. \citet{bowen2022synthetica} develops a methodology to synthesize a 207-variable dataset, which contains both categorical and continuous variables that are considered sensitive. At a high level, they first synthesize the categorical variables before applying a sequential CART model to synthesize the continuous variables. They also apply several mid-synthesis constraints, such as net capital losses may only take the values of \$0 to -\$3,000, for values that fall outside the bounds of the data. Some noise is added to the final nodes of the fitted CART model if the nodes contain very few observations. The authors use the \verb;R; package \textit{tidysynthesis} to generate the synthetic data and to assess the risk-utility trade-off. Some of the utility metrics include the first four moments, correlation fit, and CIO for regression models. The main utility metric is on an analysis-specific utility measure called tax micosimulation models, that help the public understand the potential impacts of tax policy proposals. For disclosure risk, the authors consider the number of perfectly replicated synthetic data records from confidential data, along with attribute disclosure risk using $l$-diversity \citep{machanavajjhala2007diversity}. 

\section{Concluding Remarks and Future Directions}
\label{sec:conclusion}

Thirty years of synthetic data research and applications have demonstrated huge potential of the synthetic data approach for privacy protection \citep{drechsler202330,reiter2023synthetic}. Nevertheless, as we have discussed in this article, there are still many opportunities for more researchers to take an active part. For instance, tuning risk-utility trade-off is a much under-researched area that holds promise for creating more effective and efficient synthesis. We also have not developed good synthetic data methods that handle survey weights, longitudinal, and texts well despite the great demand and abundance of such data types requiring privacy protection. Additionally, we need more computationally-feasible synthetic data approaches for spatial information. 

Turning to the synthetic data tools, we recognize the pressing need for open-source packages for synthetic data generation and evaluation. As valuable as some of the sophisticated synthetic data generation models and evaluation methods proposed in the literature may be, data curators typically will not adopt them unless they can access readily available open-source packages. This emphasizes the need for collaborative teams, including researchers, practitioners, and programmers, to develop and disseminate new synthetic data methods for their generation and evaluation in a user-friendly manner.

In addition to tools development, \citet{HuBowen2022amstat} calls for education and training opportunities for students and workforce in synthetic data (or statistical data privacy in general), which is worth restating. There have been successes in educating and training undergraduate students in synthetic data generation and evaluation for a wide range of applications (\citet{RosOlssonHu2020PSD, CaoHu2022PSD, GuoHu2023TAS} are all undergraduate student projects) and engaging undergraduates in synthetic data tools development \citep{HornbyHu2021TDP, IdentificationRiskCalculation}. We strongly advocate for more courses and training for students and workforce. By doing so, the next generation of statisticians and data scientists will be equipped with the essential knowledge, skills, and tools for using synthetic data for privacy protection. 

High profile use cases, like the 2020 Census data products in the United States, have improved communication about synthetic data and other data privacy concepts, such as differential privacy. However, the available materials are limited compared to what is needed. For instance, while many data users can likely recommend several different types of machine learning materials (e.g., books, blogs, and videos), few have recommendations for synthetic data resources \citep{snoke2020statisticians}. A similar situation exists for other arguably more popular data privacy methods, such as differential privacy and differentially private synthetic data. \citet{williams2023disclosing} surveys economists from the American Economic Association membership list who opted to receive research survey emails to assess their knowledge and attitudes regarding differential/formal privacy. One question asks if participants knew anyone within their professional circles discussed the U.S. Census Bureau's adoption of DP/formal privacy for the 2020 Decennial Census. 68.3\% report no such awareness.

These examples emphasize the need for greater involvement of statisticians and data scientists in the synthetic data and broader statistical data privacy community. As of now, most meetings and conferences tend to focus on theoretical aspects within the computer science field. Currently, there does not exist a regular conference on the intersection of data privacy and public policy. Notable recent developments indicate increasing interest in this field, such as the National Bureau of Economic Research\footnote{“Data Privacy Protection and the Conduct of Applied Research: Methods, Approaches, and their Consequences, Spring 2023,” hosted by the National Bureau of Economic Research. Accessed on June 21, 2023. https://www.nber.org/conferences/data-privacy-protection-and-conduct-applied-research-methods-approaches-and-their-consequences} and National Institute of Statistical Sciences\footnote{“IOF Workshop: Advancing Demographic Equity with Privacy Preserving Methodologies,” hosted by the National Institute of Statistical Sciences. Accessed on June 21, 2023. https://www.niss.org/events/iof-workshop-advancing-demographic-equity-privacy-preserving-methodologies} hosting workshops on data privacy and public policy. These growing needs have motivated the authors of this review to establish a conference on data privacy and public policy that is scheduled for 2024. It is our hope that more people, such as yourself, can join the community and conversation to shape the future of synthetic data and statistical data privacy.

%\section*{Acknowledgments}

\section*{Funding}
This research was partially funded by the Alfred P. Sloan Foundation grant G-2022-19513.

\section*{Contributions}
JH: Conceptualization, Writing -- original draft (lead),
and Writing -- review \& editing

\noindent CMB: Conceptualization, Writing -- original draft, and Writing -- review \& editing

\bibliographystyle{apalike}
\bibliography{ReviewBib}

\end{document}